\documentstyle[preprint,revtex]{aps}
\begin{document}
\draft
\preprint{September, 1995}
\begin{title}
Quantum duality and Bethe-Ansatz \\for the Hofstadter problem on
hexagonal lattice
\end{title}
\author{C.-A. Piguet, D.F. Wang and C. Gruber}
\begin{instit}
Institut de Physique Th\'eorique\\
Ecole Polytechnique F\'ed\'erale de Lausanne\\
PHB-Ecublens, CH-1015 Lausanne, Switzerland
\end{instit}
\begin{abstract}
The Hofstadter problem is studied on the hexagonal lattice.
We first establish a relation
between the spectra for the hexagonal lattice and for its
dual lattice, the triangular lattice. Following the idea of Faddeev and
Kashaev, we then obtain the Bethe-Ansatz equations for this system.
\end{abstract}
\narrowtext

Systems in external magnetic field have been of considerable interest
in recent years. One of the most fascinating properties of these systems is the
integer and fractional quantum Hall effect \cite{kalmeyer,pra,zou}.
The essential physics of the integer quantum Hall effect can be described
by the simple Landau problem, in which a free electron moves in a two
dimensional plane under constant magnetic field. For the fractional quantum
Hall effect, it is well known that the electron-electron correlation
gives rise to the energy gap of the system.

Besides the systems of electrons moving continuously on the two dimensional
plane
in an external magnetic field, the problem of free electrons hopping on
a two dimensional lattice under external magnetic field,
i.e. the Hofstadter problem, has attracted
a lot of attention \cite{thouless,haper,so,hof,wie,fad,lieb,zak}.
Recently, using reflection positivity, Lieb has provided a proof for the long
standing conjecture that for the square lattice, the magnetic flux which
minimises the energy of the system at half-filling is exactly $\pi$ per
plaquette \cite{lieb}. On
the other hand, in their recent work, Wiegmann and Zabrodin
have shown that the magnetic translations can be constructed with
the generators of the quantum group $U_q(sl(2))$ for the
Hofstadter problem \cite{wie}. With this representation, they obtain
the Bethe-Ansatz equations for the eigenvalue problem of the system on
a two dimensional square lattice. Later, Faddeev and Kashaev
were able to provide a generalized approach, both for the
square and triangular lattices \cite{fad}.

In two dimensions, there are three lattices of special interest:
the square, the triangular and the hexagonal lattices. In particular,
the triangular lattice and the hexagonal lattice are dual of
each other. In this paper, we study the Hofstadter problem on
the hexagonal lattice. Following the idea of
Wiegmann-Zabrodin, and Faddeev-Kashaev, we shall find
the Bethe-Ansatz type solutions for this system. Furthermore,
we establish a quantum duality relation between the energy spectra
on the triangular and on hexagonal lattices.

The general Hamiltonian for non interacting electrons
moving on a lattice in a magnetic field is:
\begin{equation}
H=\sum_{i,j}t_{ij}e^{i\theta_{ij}}c_{i}^{+}c_{j}
\label{ham}
\end{equation}
where $c_{i}^{+}$, $c_{i}$ are the creation and annihilation operators for an
electron ai site $i$, $t_{ij}$ is the real hopping matrix, and $\theta_{ij}$
corresponds to
$\int_{i}^{j}\vec{A}(\vec{x})\cdot d\vec{x}$, with $\vec{A}$ the vector
potential
and the integral is performed on a straight line between $i$ and $j$.
It thus satisfies $\theta_{ij}=-\theta_{ji}$. In the following, we consider the
special case where the electrons hop between nearest neighbours only. We
further
assume that the system is invariant under translation, but not necessarily
invariant under rotation.

The Hofstadter problem is a very interesting problem of theoretical
physics, in particular due to its distinction between rational and irrational
numbers. In the following, we only consider the rational case when the flux per
elementary
cell of the lattice is given by $\phi=2\pi\frac{M}{N}$, where $M$
and $N$ are mutually prime integers.

For the hexagonal lattice, we can define two triangular sublattices $A$ and
$B$.
Let $\vec{s}_{1}$, $\vec{s}_{2}$ and $\vec{s}_{3}$ denote the three vectors
that connect a site of the
sublattice $A$ with its three nearest neighbours of the sublattice $B$,
chosen in such a way that $\vec{s}_{1}\wedge \vec{s}_{2}$, $\vec{s}_{2}\wedge
\vec{s}_{3}$ and
$\vec{s}_{3}\wedge \vec{s}_{1}$ are in the opposite direction of the magnetic
field. We have
three hopping amplitudes $t_{1}^{h}$, $t_{2}^{h}$ and $t_{3}^{h}$ corresponding
to the three
directions defined by $\vec{s}_{1}$, $\vec{s}_{2}$ and $\vec{s}_{3}$. The index
$h$ recalls
that we have a hexagonal lattice.

Let us consider a Bloch wavefunction with vector $\vec{k}$ ($0\leq
k_{x},k_{y}\leq \frac{2\pi}{N}$):
\begin{equation}
\mid\Psi>=\sum_{\vec
n}e^{i\vec{k}\cdot\vec{n}}u_{\vec{n}}c_{\vec{n}}^{+}\mid0>,
\end{equation}
where the summation $\vec n$ is over the magnetic unit cell.
We only need to consider the above region of the momentum, as
in the rest of Brillouin zone, the situation can be mapped to
this case.

With the Hamiltonian (\ref{ham}), the Schr\"{o}dinger's equation yields the
following
equation for the coefficients $u_{\vec{n}}$ corresponding to the energy
$E^{h}$:
\begin{eqnarray}
E^{h}u_{\vec{n}}&=&
\sum_{i=1}^{3}t_{i}^{h}\alpha_{i}^{h}e^{i\theta_{\vec{n},\vec{n}+\vec{s}_{i}}}u_{\vec{n}
+\vec{s}_{i}}\,\,\,\,{\rm if}\,\,\vec{n}\in A
\label{ua}\\
E^{h}u_{\vec{n}}&=&
\sum_{i=1}^{3}t_{i}^{h}(\alpha_{i}^{h})^{-1}e^{i\theta_{\vec{n},\vec{n}-\vec{s}_{i}}}
u_{\vec{n}-\vec{s}_{i}}\,\,\,\,{\rm if}\,\,\vec{n}\in B
\label{ub}
\end{eqnarray}
where $\alpha_{i}^{h}=e^{i\vec{k}\cdot\vec{s}_{i}}$.
Combining (\ref{ub}) and (\ref{ua}), we obtain the eigenvalue equation for the
coefficients $u_{\vec{n}}$ of the sublattice $A$:
\begin{eqnarray}
&
&\left\{(E^{h})^{2}-((t_{1}^{h})^{2}+(t_{2}^{h})^{2}+(t_{3}^{h})^{2})\right\}u_{\vec{n}}
=\nonumber\\
& &\sum_{i\neq j=1}^{3}t_{i}^{h}t_{j}^{h}\alpha_{i}^{h}(\alpha_{j}^{h})^{-1}
e^{i\theta_{\vec{n},\vec{n}+\vec{s}_{i}}+i\theta_{\vec{n}+\vec{s}_{i},\vec{n}+\vec{s}_{i}-\vec{s}_{j}}}
u_{\vec{n}+\vec{s}_{i}-\vec{s}_{j}}.
\end{eqnarray}

At this point, we define the three vectors
$\vec{S}_{1}=\vec{s}_{2}-\vec{s}_{3}$,
$\vec{S}_{2}=\vec{s}_{3}-\vec{s}_{1}$ and $\vec{S}_{3}=\vec{s}_{1}-\vec{s}_{2}$
which connect nearest neighbours of the sublattice $A$. Using the fact that:
\begin{equation}
\theta_{\vec{n},\vec{n}+\vec{s}_{i}}+\theta_{\vec{n}+\vec{s}_{i},\vec{n}+\vec{s}_{i}-\vec{s}_{j}}
=\epsilon_{ijk}\frac{\phi}{6}+\theta_{\vec{n},\vec{n}+\epsilon_{ijk}\vec{S}_{k}}
\end{equation}
with $\epsilon_{ijk}$ the Levi-Civita symbol and $\phi$ the flux through
elementary hexagons, we have:
\begin{eqnarray}
&
&\left\{(E^{h})^{2}-((t_{1}^{h})^{2}+(t_{2}^{h})^{2}+(t_{3}^{h})^{2})\right\}u_{\vec{n}}
=\nonumber\\
& &\sum_{i\ne j\ne k=1}^3 t_{i}^{h}t_{j}^{h}\alpha_{i}^{h}(\alpha_{j}^{h})^{-1}
\omega^{\epsilon_{ijk}\frac{1}{6}}
e^{i\theta_{\vec{n},\vec{n}+\epsilon_{ijk}\vec{S}_{k}}}u_{\vec{n}+\epsilon_{ijk}\vec{S}_{k}},
\end{eqnarray}
where $\omega=e^{i\phi}$.

Let us then consider the Hofstadter problem on the triangular lattice. With
$t_{1}^{t},t_{2}^{t},t_{3}^{t}$ the three hopping amplitudes the
Schr\"{o}dinger's equation gives the following equation for the coefficients
$u_{\vec{n}}$ corresponding to the energy $E^{t}$:
\begin{equation}
E^{t}u_{\vec{n}}=
\sum_{i=1}^{3}\left(t_{i}^{t}\alpha_{i}^{t}e^{i\theta_{\vec{n},\vec{n}+\vec{S}_{i}}}u_{\vec{n}+\vec{S}_{i}}
+t_{i}^{t}(\alpha_{i}^{t})^{-1}e^{i\theta_{\vec{n},\vec{n}-\vec{S}_{i}}}u_{\vec{n}-\vec{S}_{i}}\right )
\end{equation}
where $\alpha_{i}^{t}=e^{i\vec{k}\cdot\vec{S}_{i}}$.

With the above relations,
one can now easily relate the energies of the hexagonal lattice to the ones of
the
triangular lattice:
\begin{equation}
(E^{h})^{2}-((t_{1}^{h})^{2}+(t_{2}^{h})^{2}+(t_{3}^{h})^{2})=E^{t}
\label{energy}
\end{equation}
where we have the following correspondances
of the hopping parameters and of the momenta:
\begin{eqnarray}
t_{1}^{t}&=&t_{2}^{h}t_{3}^{h}\nonumber\\
t_{2}^{t}&=&t_{3}^{h}t_{1}^{h}\nonumber\\
t_{3}^{t}&=&t_{1}^{h}t_{2}^{h}\nonumber\\
\alpha_{1}^{t}&=&\alpha_{2}^{h}(\alpha_{3}^{h})^{-1}\omega^{\frac{1}{6}}\nonumber\\
\alpha_{2}^{t}&=&\alpha_{3}^{h}(\alpha_{1}^{h})^{-1}\omega^{\frac{1}{6}}\nonumber\\
\alpha_{3}^{t}&=&\alpha_{1}^{h}(\alpha_{2}^{h})^{-1}\omega^{\frac{1}{6}}.\nonumber\\
\label{cor}
\end{eqnarray}
It is well known that the spectrum of any bipartite lattice is symmetric around
zero.
This can be understood from the fact that the Hamiltonian
$H$ can be transformed in $-H$ through the following unitary transformation:
\begin{equation}
c_{x}\longrightarrow \left\{\begin{array}{ccc}
c_{x} & {\rm if } & x\in A \\
-c_{x} & {\rm if } & x\in B. \\
\end{array}\right.
\end{equation}
The hexagonal lattice is bipartite, while this is not the case for the
triangular lattice. Therefore the spectrum on the hexagonal lattice is
symmetric
around zero and this property does not hold for the triangular lattice.
This result is clearly seen with Eq.(\ref{energy}).

The duality relation expressed by Eq.(\ref{energy}) and Eq.(\ref{cor}) allow us
to use directly the results of
Faddeev and Kashaev \cite{fad} to find the Bethe-Ansatz equations for the
hexagonal system.
Let us recall briefly the results for the triangular lattice.
Using a $N^3$ reducible representation, the Hilbert space is a tensor product
of
three subspaces, each of which has dimension $N$.
A wavefunction can thus be written as
\begin{equation}
\mid\phi>=\mid\phi>_{0}\otimes\mid\phi>_{1}\otimes\mid\phi>_{2}.
\end{equation}
Then, introduce the following notation:
\begin{eqnarray}
&&\alpha_1^t=e_2^{1/2} (f_2 C)^{-1/2}, \alpha_2^t=e_1^{1/2}
(f_1 C)^{-1/2}, \alpha_3^t=e_0^{1/2} (f_0 C)^{-1/2},\nonumber\\
&&t_1^t=(e_2f_2 C)^{1/2}, t_2^t=(e_1f_1 C)^{1/2}, t_3^t= (e_0 f_0 C)^{1/2}.
\label{relation}
\end{eqnarray}
One also has
\begin{equation}
e_i=b_{i-1} d_i c_{i+1}, f_i=c_{i-1} a_i b_{i+1},
\end{equation}
with $i \in Z_{3}$.
Here $C$ is a constant complex number, $C^N=(-1)^{(N-1)}$.

The eigenvalue equation for the projection $Q(p)$ of an eigenvector $\mid\phi>$
on the Baxter's vector
$\mid p>$, which corresponds to the point $p=(x,\xi_0,\xi_1,\xi_2)$ on a
curve in a four dimensional space, reads
\begin{equation}
\Lambda(x)Q(p)=Q(\tau_{-}p)\Delta_{-}(p)+Q(\tau_{+}p)\Delta_{+}(p), \,\,
Q(p)=<\phi|p>,
\end{equation}
where $\Lambda(x)$ is related to the eigenvalue $E^t$ by
\begin{equation}
\Lambda(x)=\omega a_{0}a_{1}a_{2}C+d_{0}d_{1}d_{2}+x^{2}E^t,
\end{equation}
$\tau_{+}$ and $\tau_{-}$ are defined on the coordinates
$x,\xi_{0},\xi_{1},\xi_{2}$ by
\begin{equation}
\tau_{\pm}x=\omega^{\pm1/2}x,\,\,\tau_{\pm}\xi_{i}=\omega^{-1/2}\xi_{i},
\,\,i=0,1,2,
\end{equation}
and
\begin{eqnarray}
&&\Delta_{-}(p)=\prod_{i \in Z_{3}}(d_{i}-x\xi_{i+1}c_{i})\nonumber\\
&&\Delta_{+}(p)=\prod_{i \in
Z_{3}}\xi_{i}(a_{i}d_{i}-x^{2}b_{i}c_{i})/(\xi_{i+1}a_{i}-xb_{i}).
\end{eqnarray}
The zeros $p_{k}$ of $Q(p)$ are given by the following Behte-Ansatz equation
\begin{equation}
\frac{Q(\tau_{-}p_{k})}{Q(\tau_{+}p_{k})}=
-\frac{\Delta_{+}(p_{k})}{\Delta_{-}(p_{k})}.
\end{equation}
The equations for the hexagonal lattice take the same form except that one has
to
introduce $(\ref{cor})$ into the definitions $(\ref{relation})$.


In the special case where the genus of the curve
vanishes and $N$ is odd ($N=2P+1$), one may simplify the above equations, so
that they
can be written out explicitly. This special case corresponds
to the values of momenta: $\alpha_1^t=\alpha_2^t=\alpha_3^t=q^{1/2}$, with
$q=\omega^{1/2}$.
The energy for the hexagonal lattice reads:
\begin{eqnarray}
(E^{h})^{2}&=&(t_{1}^{h})^{2}+(t_{2}^{h})^{2}+(t_{3}^{h})^{2}+
(q^{1/2}+q^{-1/2})(t_{2}^{h}t_{3}^{h}+t_{3}^{h}t_{1}^{h}+t_{1}^{h}t_{2}^{h})\nonumber\\
&
&-(q-q^{-1})((t_{1}^{h})^{2}t_{2}^{h}t_{3}^{h}+t_{1}^{h}(t_{2}^{h})^{2}t_{3}^{h}
+t_{1}^{h}t_{2}^{h}(t_{3}^{h})^{2})\sum_{m=1}^{2P}z_{m}\nonumber\\
& &+(t_{1}^{h}t_{2}^{h}t_{3}^{h})^{2}(q-q^{-1})(q^{1/2}-q^{-1/2})\sum_{1\leq
m<n\leq2P}z_{m}z_{n}
\end{eqnarray}
where the $z_{l}$ ($l=1,\ldots,2P$) are given by the Bethe-Ansatz equations:
\begin{equation}
\prod_{m=1,m\neq l}^{2P}\frac{qz_{l}-z_{m}}{z_{l}-qz_{m}}=
q^{-1/2}\frac{(t_{2}^{h}t_{3}^{h}z_{l}+q^{1/2})(t_{3}^{h}t_{1}^{h}z_{l}+q^{1/2})
(t_{1}^{h}t_{2}^{h}z_{l}+q^{1/2})}{(q^{1/2}t_{2}^{h}t_{3}^{h}z_{l}-1)
(q^{1/2}t_{3}^{h}t_{1}^{h}z_{l}-1)(q^{1/2}t_{1}^{h}t_{2}^{h}z_{l}-1)}.
\end{equation}

In summary, we have developed the Bethe-Ansatz for the Hofstadter problem on
the hexagonal lattice in this paper. An interesting duality is discovered
between the hexagonal lattice and its dual partner (triangular lattice).
Using this duality relation, we have written the Behte-Ansatz equations for the
point of the magnetic Brillouin zone where the equations of Faddeev and Kashaev
take an explicit form.
Further work is necessary to obtain more information from these equations.

We wish to thank Prof. H. Kunz, Prof. F. Reuse,
Prof. S. Maumary, Prof. Y. S. Wu, Prof. Mo-lin GE and Prof. E. H. Lieb
for conversations. This work was supported in part
by the Swiss National Science Foundation.


\end{document}